\pdfoutput=1 \pdfsuppresswarningpagegroup=1 
\documentclass[
 aps,%
 prl,%
 reprint, twocolumn,%
 groupedaddress,% 
 superscriptaddress,%
  showpacs,%
 amsfonts,%
 letterpaper,%
 footinbib,%
 noeprint%o
]{revtex4-1}

\RequirePackage{amsmath,amssymb,bm}
\RequirePackage{graphicx}
\RequirePackage[usenames,dvipsnames]{xcolor}

\usepackage{hyperref}
\hypersetup{
    colorlinks=true,
    citecolor=blue,
    linkcolor=magenta,
    filecolor=cyan,      
    urlcolor=magenta,
    pdfauthor = {Chuan Chen},
    pdfcreator = {\LaTeX\ and \flqq hyperref\frqq},
}

\usepackage{pdfpages}
\makeatletter\AtBeginDocument{\let\LS@rot\@undefined}\makeatother

\graphicspath{ {./Figures/} } 

\newcommand{\Fref}[1]{Fig.~\ref{#1}}
\newcommand{\Frefs}[1]{Figs.~\ref{#1}}
\newcommand{\Eqref}[1]{Eq.~\eqref{#1}}

\newcommand{\SFref}[1]{supplementary Fig.~S{#1}}
\newcommand{\SEqref}[1]{supplementary Eq.~S{#1}}

\newcommand{\br}{\mathbf{r}}
\newcommand{\bq}{\mathbf{q}}
\newcommand{\bQ}{\mathbf{Q}}
\newcommand{\bG}{\mathbf{G}}

\newcommand{\bnabla}{\bm{\nabla}}

\newcommand{\TS}{TiSe\textsubscript{2}}

\newcommand{\tapprox}{{\,\approx\,}}

\newcommand{\tequiv}{{\,\equiv\,}}
\newcommand{\teq}{{\,=\,}}

\newcommand{\tsimeq}{{\,\simeq\,}}
\newcommand{\tsim}{{\,\sim\,}}
\newcommand{\tlt}{{\,<\,}}
\newcommand{\tle}{{\,\le\,}}
\newcommand{\tgt}{{\,>\,}}

\newcommand{\tminus}{{\,-\,}}
\newcommand{\tplus}{{\,+\,}}

\newcommand{\ttimes}{{\times}}
\newcommand{\tpropto}{{\,\propto\,}}
\newcommand{\tto}{{\,\to\,}}

\begin{document}

\title{Discommensuration-driven superconductivity in the charge density wave 
phases of transition-metal dichalcogenides}

\author{Chuan Chen}
\affiliation{Centre for Advanced 2D Materials and Graphene Research Centre, 
National University of Singapore, Singapore 117546}  
\affiliation{Department of Physics, National University of Singapore, Singapore 
117542} 

\author{Lei Su}
\affiliation{Department of Physics, University of Chicago, Chicago, Illinois 60637, USA}  

\author{A.~H. {Castro Neto}}
\affiliation{Centre for Advanced 2D Materials and Graphene Research Centre, 
National University of Singapore, Singapore 117546}  
\affiliation{Department of Physics, National University of Singapore, Singapore 
117542} 

\author{Vitor M. Pereira}
\affiliation{Centre for Advanced 2D Materials and Graphene Research Centre, 
National University of Singapore, Singapore 117546}  
\affiliation{Department of Physics, National University of Singapore, Singapore 
117542} 


\date{\today}

\begin{abstract}
We introduce a McMillan-Ginzburg-Landau theory to describe the cooperative 
coexistence of charge-density and superconducting order in two-dimensional 
crystals. 
With a free-energy that explicitly accounts for the competition between 
commensurate and incommensurate ground states, we are able to map the transition 
between these phases and monitor the development of discommensurations in the 
near-commensurate regime. 
Attributing the enhancement of superconducting order to density-wave 
fluctuations, we propose a coupling scheme that yields a phase diagram 
in qualitative agreement with experiments in conducting transition metal 
dichalcogenides. 
The model predicts the development of non-uniform superconductivity 
similar to that arising from a pair-density wave, with a spatial texture  
driven by the underlying charge-density wave fluctuations.
\end{abstract}

\maketitle

% ------------------------------------------------------------------------------
% MAIN TEXT BEGINS
% ------------------------------------------------------------------------------

Recent experiments suggest a relation between emergent superconductivity in 
doped transition-metal dichalcogenides (TMDs) and fluctuations of their charge 
density wave (CDW) order \cite{Li2015,Yan2017,Novello2017,Wang2017}.
The archetype example of 1T-TiSe$_2$ (\TS{} in short) displays 
superconductivity (SC) amidst CDW order as soon as the nature of the 
latter changes from commensurate (C) to incommensurate (IC) under electron 
doping \cite{Morosan2006,Joe2014,Li2015,Kogar2017} or pressure 
\cite{Kusmartseva2009,Joe2014}, either in bulk or 2D samples 
\cite{Li2015,Li2018}. 
SC is limited to a dome over a small range of the external 
parameter $x$ (doping or pressure) in the $T$--$x$ phase diagram. Since 
CDW correlations persist in the SC phase \cite{Spera2017} and the dome is 
centered at the putative quantum critical point of the commensurate CDW (C-CDW) 
phase, it has been suggested that SC might arise (or be enhanced) as a result of 
CDW fluctuations \cite{Barath2008,Wang2017,Chen:2017}. 

The basic excitation of a C-CDW is called \emph{discommensuration} (DC)
\cite{McMillan1976}: a localized defect (domain wall) where the phase of 
the order parameter jumps by $2\pi\nu$, with $\nu$ the commensurability 
fraction \cite{McMillan1976,Bak1976,GrunerBook}. 
DCs have been observed in \TS{} by STM \cite{Yan2017,Novello2017} above the 
optimal SC transition temperature ($T_\text{sc}^\text{max} \tsimeq 4$\,K), and 
are implied by inelastic scattering \cite{Kogar2017}. This suggests that the CDW 
converts from C to IC through a near-commensurate (NC) regime characterized by a 
finite density of DCs, similarly to the cases of 2H-TaSe$_2$ \cite{McMillan1976} 
or 1T-TaS$_2$ \cite{Thomson1994a}. 

Although the range $T \tlt T_\text{sc}$ remains unexplored by STM, Little-Parks 
magnetoresistance oscillations \cite{Little1962} observed in \TS{} films 
\cite{Li2015} were interpreted as a result of supercurrents constrained 
by an underlying periodicity tied to the CDW background. 
STM observations of enhanced density of states within DCs \cite{Yan2017} 
indirectly support this. Moreover, the onset of a DC network introduces new 
low-energy phonons \cite{McMillan1977,Nakanishi1978a} that can couple to 
electrons and induce a Cooper instability \cite{SI}. 
Both ingredients suggest that the underlying theory must tie SC to 
both fluctuations and the domain structure of the electronic CDW.

To investigate the potential role of CDW fluctuations in either inducing or 
enhancing the SC order, we propose an extension of McMillan's Ginzburg-Landau 
framework for the CDW in layered TMDs \cite{McMillan1975,McMillan1976}. It 
incorporates a SC order parameter coupled to the electronic density via DCs. In 
the vicinity of the C-IC transition (the NC regime), the predicted phase diagram 
reproduces the experimental one in \TS{} with no fine-tuning of parameters (all 
$\sim 1$). 
The nature of the SC phase is interesting and novel: the model implies 
non-uniformity in the NC regime close to $T_\text{sc}$ and, with decreasing 
temperature, SC order might sequentially percolate from 0d to 1d to 2d.

\textbf{CDW order}\;---\;%
McMillan established the approach to the C-IC transition in terms of a free 
energy functional with a complex order parameter 
\cite{McMillan1975,McMillan1976}. 
Although the approach is general, the relevant nonlinear and \emph{umklapp} 
terms depend on the particular ordering vectors and commensurability 
condition \cite{McMillan1975}.
To be specific, we consider here the case of \TS{} since its small carrier 
density makes it an easily tuneable system \cite{Joe2014,Kogar2017,Li2015}.
Both bulk \cite{Salvo1976} and monolayer \cite{Chen2015} \TS{} undergo a second 
order phase transition to the C-CDW phase characterized by the formation of a 
$2\ttimes2$ superlattice in the 2d planes. 
The experimentally measured density modulation $\delta\rho(\br)$ 
is contributed by three plane waves with wavevectors $\bQ_j^C \tequiv 
\bG_j/2$, where $\bG_j$ ($j \teq 1,2,3$) are primitive reciprocal vectors 
related by $C_3$ rotations \cite{Salvo1976}.
As the in-plane ordering is the same in both bulk and monolayer \cite{Chen2015}, 
we neglect the inter-layer coupling and focus on the doping--temperature 
phase diagram of a \TS{} monolayer \cite{Kogar2017,Li2015}. We ignore electronic 
disorder \cite{note-disorder}, as appropriate for gate-induced doping in 
encapsulated few-layer systems \cite{Li2015}, or doping by Cu intercalation that 
donates conduction electrons without visible disruption of the electronic 
bandstructure \cite{Morosan2006,Chen:2017}.
Following the approach of references 
\cite{McMillan1975,Jacobs1980,Nakanishi1977}, we define the complex CDW order 
parameters, $\psi_j(\br) \tequiv \varphi_j(\br) e^{i\theta_j(\br)}$, according 
to 
\begin{equation}
  \delta\rho(\br) \tequiv \textstyle\sum_j
    e^{i \br\cdot\bQ_j^C} \psi_j (\br) + \text{c.c.},
  \label{eq:def-psi}
\end{equation}
where $\psi_j(\br)$ encodes deviations from the C state.
To describe the IC phase, we introduce the wavevectors $\bQ^I_j$ that 
parametrize a uniform IC-CDW with the same symmetry. In line with 
experiments \cite{SI}, we take $\bQ^I_j \teq (1+\delta)\bQ^C_j$, where $\delta$ 
quantifies the incommensurability, and further define $\bq_{j}^I \tequiv 
\bQ_j^I-\bQ^C_j$, $q^I \tequiv |\bq_j^I| \teq  \delta\,|\bQ^C_j|$.

The free energy density consists of a conventional Ginzburg-Landau portion,
\begin{subequations} \label{eq:f-cdw}
\begin{equation}
  f_0 (\br) \equiv 
  A \sum_j |\psi_{j}|^2 
  + B\sum_{j} \bigl| (i\bnabla_j + \bq_{j}^I) \, \psi_{j} \bigr|^2 
  + G\sum_{j}| \psi_{j}|^4 
  ,
  \label{eq:f0}
\end{equation}
where the $B$ term favors a solution $\psi_j(\br) \tpropto 
e^{i\bq_j^I\cdot\br}$ that distorts $\delta\rho(\br)$ towards an IC state
\footnote{The actual magnitude of $\bq_j^I$ does not play a role in the 
subsequent energy minimization because it can be absorbed into the definition 
of $B$.}.
The quadratic coefficient is assumed to vanish linearly at a critical 
temperature: $A \tequiv t \propto T \tminus T_\text{icdw} $,
$t$ being the reduced temperature.
The presence of non-colinear waves contributing to $\delta\rho(\br)$ entails 
additional terms in the free energy to 4th order \cite{McMillan1975,Jacobs1980}. 
Symmetry dictates them to be \cite{SI}
\begin{multline}
  f_1(\br) \equiv 
  - \frac{E}{2} \sum_{j}\left(\psi_{j}^{2}+\psi_{j}^{*2}\right) 
  - \frac{3D}{2}\left(\psi_1\psi_2\psi_3 + \text{c.c.} \right)
  \\
  - \frac{M}{2} \sum_{j}\left(\psi_j\psi_{j+1}^{*}\psi_{j+2}^{*} + \text{c.c.}
    \right)
  + \frac{K}{2}\sum_{i\ne j} |\psi_i\psi_j|^2
  .
  \label{eq:f1}
\end{multline}
\end{subequations}
The total CDW free energy reads $\mathcal{F}_\text{cdw} \tequiv \int 
[f_0(\br) + f_1(\br) ] \, d\br$. The subscript $j$ runs cyclically over 
$\{1,2,3\}$ in all our expressions (e.g., $\psi_5\tequiv\psi_2$).
Physically, the last 3 terms in \Eqref{eq:f1} reflect the electrostatic cost 
incurred by the superposition of distinct density waves
\footnote{The form of interactions between density waves here is specific to the 
case of \TS{}.}.
The $E$ term represents the \emph{lock-in} energy since it lowers the total 
energy of a C-CDW but averages out for an IC-CDW, thereby favoring the former.

\Eqref{eq:f1} induces harmonics of any IC-CDW characterized by 
$\psi_j\propto e^{i\bq_j\cdot\br}$, implying that the equilibrium IC state 
consists of a linear combination of all compatible harmonics and making the 
analytical minimization of $\mathcal{F}_\text{cdw}$ a formidable task. We tackle 
the problem numerically with a systematic expansion 
of the order parameter, as pioneered  by Nakanishi \emph{et al.} 
\cite{Nakanishi1977a,Nakanishi1977,Nakanishi1978}. The method amounts to 
expanding each $\psi_j(\br)$ in terms of $e^{i\bq_j\cdot\br}$ 
and all the two-dimensional harmonics spawned by the nonlinear terms in 
\Eqref{eq:f1} \cite{SI}. This converts $\mathcal{F}_\text{cdw}$ 
from a functional of $\psi_j(\br)$ into a function of a countable set of 
amplitudes $\Delta_{j;lmn}$ and wavevectors $\bq_{j;lmn}$ of the different 
harmonics. The equilibrium solution follows from minimizing 
$\mathcal{F}_\text{cdw}$ with respect to these parameters as well as $\bq_j$ 
itself. We take $\bq_j {\,\parallel\,} \bq_j^I$, and introduce $\eta \tequiv 
|\bq_j| / q^I$ that determines if the solution is a C-CDW ($\eta\teq 0$), a 
uniformly IC-CDW ($\eta\teq 1$), or in between (NC-CDW).

% ------------------------------------------------------------------------------
% FIGURE
% ------------------------------------------------------------------------------
\begin{figure}
\centering
\includegraphics[width=0.45\textwidth]{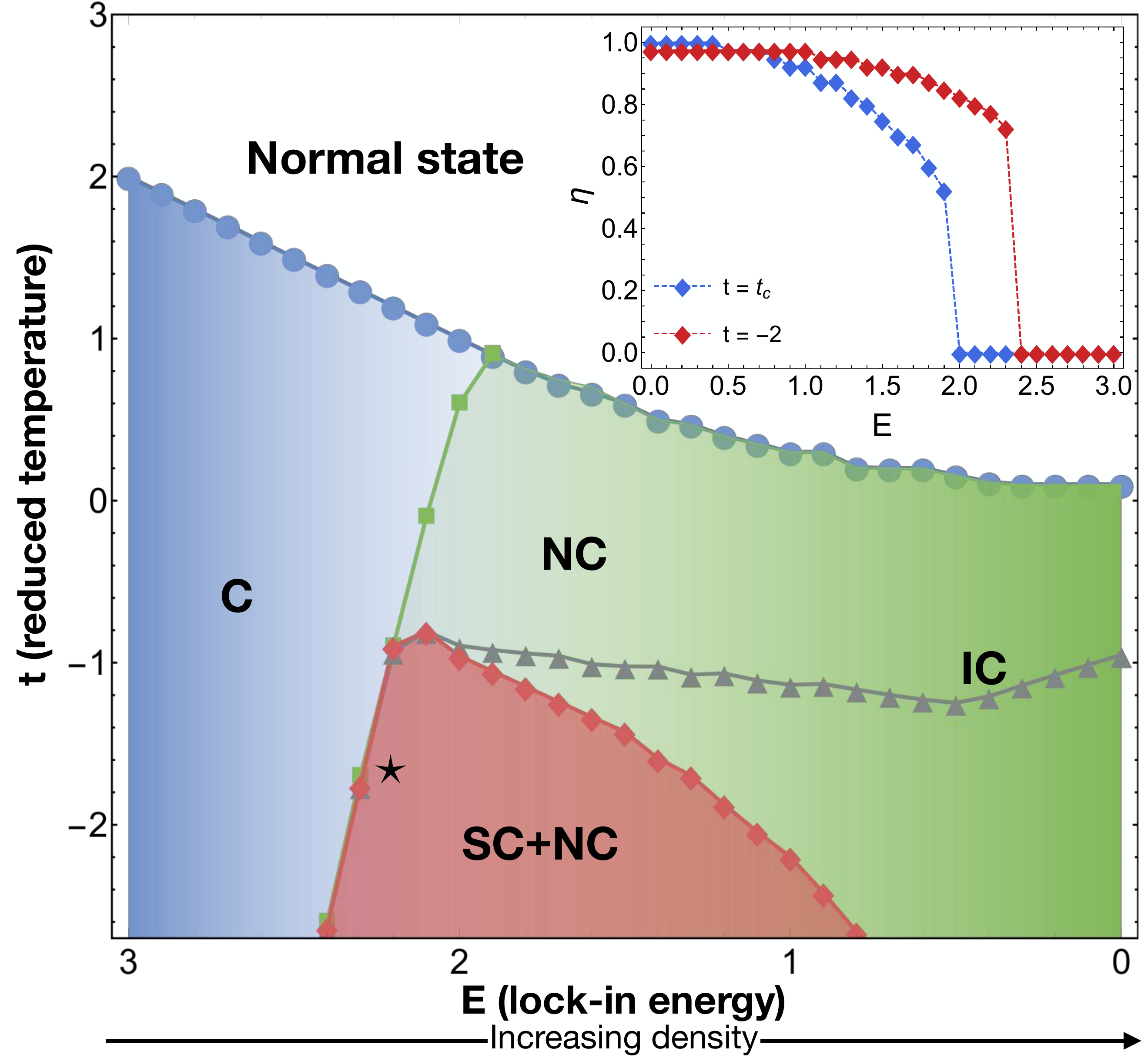}
\caption{
Phase diagram obtained by minimizing $\mathcal{F}_\text{cdw}$.
Labels C, NC and IC stand for commensurate, near-commensurate and homogeneously 
incommensurate CDW phases, respectively.
When $\mathcal{F}_{\text{cdw}} \tlt 0$, the system is in a CDW state and
the C phase corresponds to $\eta \teq 0$.
The green line represents the C-IC boundary, $E_c(t)$.
The red line indicates the boundary of the SC phase including the linear $E$ 
dependence in the CDW-SC coupling $a_s$ of \Eqref{eq:f-sc} ($a_1 = 500 E$);
it becomes the gray line if $a_s$ is $E$-independent ($a_1 = 500\ttimes 2.1$).
The inset shows the equilibrium $\eta$ at 
$t_c$ (first order transition)
%(the transition is first order) 
and at low temperature.
}
\label{fig:phase-diagram}
\end{figure}
% ------------------------------------------------------------------------------

\textbf{CDW phase diagram}\;---\;%
As we are only interested in scrutinizing the C-IC transition, we map the phase 
diagram in the $E$--$t$ plane fixing the remaining parameters to $A \teq t$, $K 
\teq G \teq 2 B (q^I)^2 \teq {-}2D \teq 2M \teq 2$ \cite{SI}. Without any fine 
tuning, this choice already allows us to concentrate on the C-IC boundary shown 
in \Fref{fig:phase-diagram} and drive the transition via $E$, which controls 
the energy gain of having a C-CDW. 
Physically, a \emph{smaller} $E$ can be mapped to \emph{larger} electron 
densities because: (i) phenomenologically, electron doping reduces the stability 
of the C state in favor of an IC one \cite{Li2015,Kogar2017,Yan2017}; (ii) the 
lock-in gain reflects the condensation energy of the C-CDW phase in a 
microscopic description, and the latter has been shown to decrease 
with doping in the excitonic theory for the C-CDW in \TS\ 
\cite{Lee1974,Kotani1977,Kotani1977a,SI}.
For this reason, the horizontal $E$ axis in the figure is reversed so that 
electron densities increase from left to right.

The phase diagram in \Fref{fig:phase-diagram} exhibits the anticipated stability 
of the C state at large $E$ (low density) and its suppression below a critical, 
temperature-dependent lock-in parameter: $E_c(t)$. Note that the critical 
temperature, $t_c(E)$, decreases when progressing from the C to the IC 
state, in agreement with the experimental trend \cite{Kogar2017,Li2015}. 
Likewise in agreement is the abrupt loss of the C phase indicated by the 
steep slope of the line $E_c(t)$.
In light of our earlier definition of $t$, the asymptotic tendency $t_c 
(E{\,\to\,}0) \tapprox 0$ means that $T_c {\,\to\,} T_\text{icdw}$, or that, as 
expected from \eqref{eq:f0}, a uniform IC state is ultimately preferred in 
the absence of lock-in energy. The inset shows the equilibrium value of $\eta$ 
at the critical temperature of the normal--CDW transition and at low 
temperatures: It grows towards $\eta \tapprox 1$ with decreasing $E$, implying 
that the dominant wavevectors contributing to $\delta\rho(\br)$ increasingly 
approach the reference IC vector $\bQ^I_j$. 

\begin{figure}
\centering
\includegraphics[width=0.5\textwidth ]{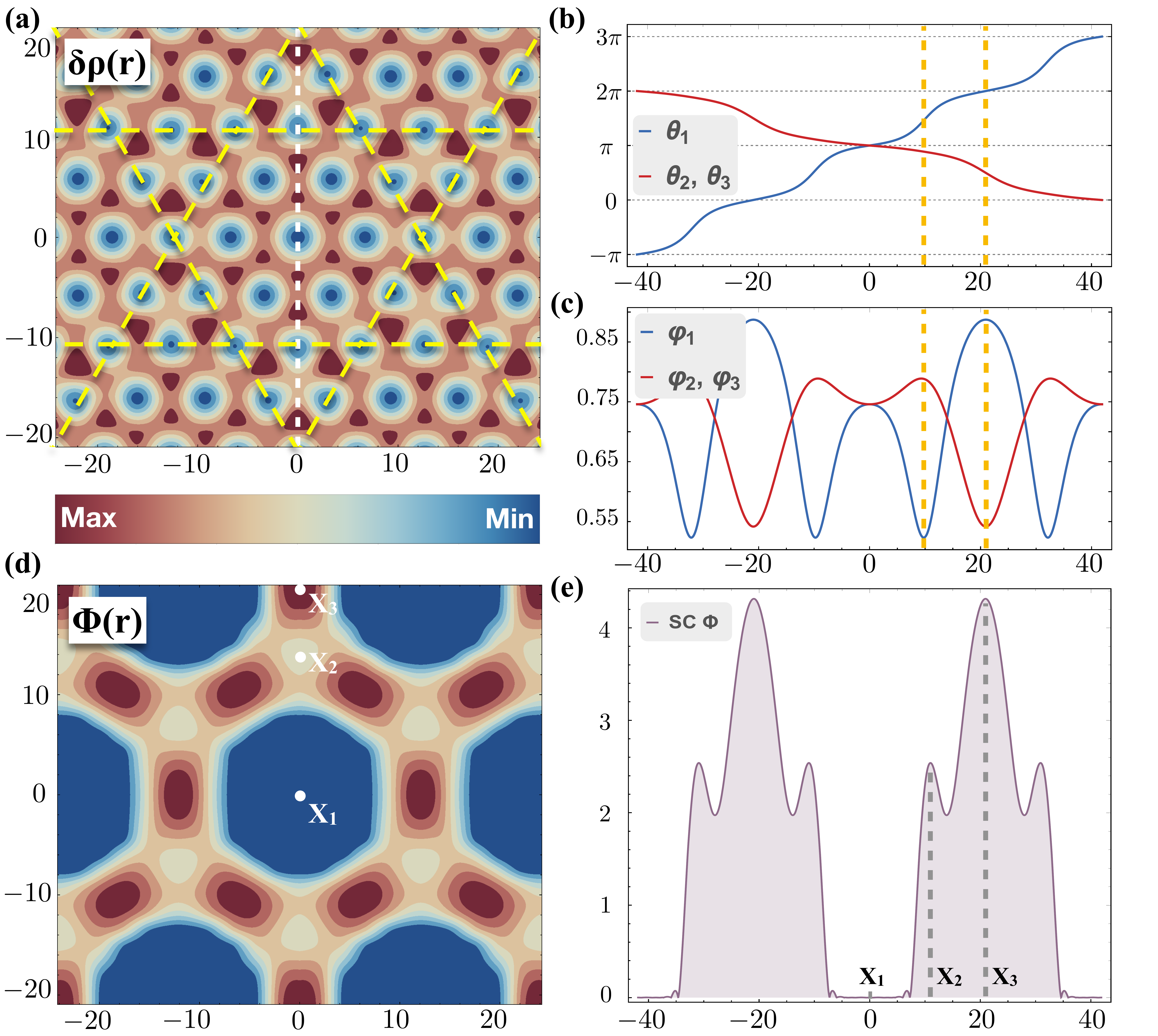}
\caption{
(a) Real space plot of the density profile $\delta\rho(\br)$ at $E \teq 2.2$,
$t \teq {-}1.7$ (in units of $\sqrt{3}a/2\pi$, with $a$ the lattice constant).
The yellow-dashed lines mark the places where the phase of each CDW order 
parameter, $\psi_j(\br)$, jumps by $\pi$. 
(b) and (c) respectively show the phase and amplitude of $\psi_j(\br)$ along the 
white vertical cut marked in (a).  
(d) The SC order parameter, $\Phi(\br)$, in the same region as (a). 
(e) $\Phi(\br)$ along the vertical cut marked in (a).   
}
\label{fig:real-space}
\end{figure}

Knowledge of $\eta$ is insufficient to characterize the rich spatial texture of 
the charge modulation which depends on the detailed harmonic content
that minimizes $\mathcal{F}_\text{cdw}$ (\SEqref{6}). 
\Fref{fig:real-space}(a) shows $\delta\rho(\br)$ at the representative point 
close to the C-CDW boundary marked by $\star$ in \Fref{fig:phase-diagram}.
\Frefs{fig:real-space}(b,c) show line cuts of the phase and amplitude of 
the order parameters $\psi_j(\br) \tequiv \varphi_j(\br)e^{i\theta_j(\br)}$ 
along the vertical dashed line in panel (a).
The phase $\theta_j(\br)$ displays a stepwise variation with 
periodic slips of $\pi$. Since \eqref{eq:def-psi} implies that regions 
where $\theta_j(\br) \tapprox 0\!\!\mod\pi$ are commensurate with the Bravais 
lattice, the spatial profile of the phase reveals an equilibrium state 
characterized by domains of approximately C-CDW separated by DCs of $\pi$.
This NC regime replicates the characteristics of CDW domain walls 
investigated by STM slightly above $T_\text{sc}$ in \TS{} 
\cite{Novello2017,Yan2017}.

Adapting \Eqref{eq:f1} to a general commensurability 
condition $\bQ^C \teq \nu \bG$ with $\nu$ a rational number ($\nu\teq 1/2$ for 
\TS), one obtains a corresponding domain structure with phase steps of $2\pi\nu$ 
across domain boundaries \cite{Bak1976, McMillan1977, Nakanishi1977, 
Nakanishi1978, Jacobs1980}. 
In 1d phase-only reductions of this problem [$\varphi_j(\br) 
\teq\text{const.}$], the saddle-point condition for $\mathcal{F}_\text{cdw}$ 
becomes a sine-Gordon equation \cite{Bak1976,McMillan1977} and DCs correspond 
to its soliton solutions. 
Even though our problem of interest is two-dimensional, 
\Eqref{eq:def-psi} still consists of a linear combination of 1d CDW modulations 
along each $\bG_j$. It is thus not surprising that each $\theta_j(\br)$ in 
\Fref{fig:real-space}(b) retains a soliton-like nature.

The DCs form a 2D Kagome superlattice overlaying the C-CDW, as highlighted by 
the yellow-dashed contours in \Fref{fig:real-space}(a). For a general 
commensurability fraction $\nu$, the period of the DC network is $L \teq 
2\pi\nu / (\eta q^I) \teq \sqrt{3} a / (\eta\delta)$, where $a$ is the 
lattice constant of the crystal in the normal phase.

Note that the amplitude of $\psi_i(\br)$ is also significantly modulated: 
\Fref{fig:real-space}(c) shows it can drop more than 30\% at each DC. The high 
variational freedom possible in our harmonic expansion permits the CDW to 
distort in order to minimize both the lock-in and gradient terms of 
$\mathcal{F}_\text{cdw}$.
The solution thus acquires both C \emph{and} IC features, consisting of domains 
with nearly flat phase and high amplitude (C-CDW), joined by domain boundaries 
where the amplitude drops to lessen the cost in deviating from 
commensurability, and the phase jumps so that, on spatial average, 
$\langle\theta_j(\br)\rangle \tapprox \bq^I_j\cdot\br$ (IC-CDW).

% ------------------------------------------------------------------------------
% FIGURE
% ------------------------------------------------------------------------------
% ------------------------------------------------------------------------------

\textbf{Coupling to superconductivity}\;---\;%
It is natural to expect these DCs to couple strongly with the SC order 
parameter: On the one hand, the development of a DC superlattice as in 
\Fref{fig:real-space}(a) introduces new low energy phonons \cite{McMillan1977, 
Nakanishi1978a,SI} that might enhance any intrinsic phonon-mediated pairing 
tendency.
On the other hand, DCs are but CDW fluctuations. While both phase and 
amplitude fluctuations are gapped in the C regime \cite{GrunerBook}, the 
transition to the NC state releases them to potentially favor SC through 
fluctuation-induced pairing.

As a minimal approach to describe the interplay between the two orders, we 
propose extending the conventional \cite{Tinkham2004} Ginzburg-Landau free 
energy associated with the SC order parameter, $\Phi(\br)$, by writing
\begin{equation}
  \mathcal{F}_\text{sc} \equiv \int
  \Bigl[ 
    a_s(T, \bnabla\psi_j) \, |\Phi|^2 
    + b_s |\bnabla\Phi|^2 
    + c_s |\Phi|^4
  \Bigr] d\br
  \label{eq:f-sc} .
\end{equation}
Making $a_s$ a function of $\bnabla\psi_j$ permits the enhancement of SC by 
deviations from a C-CDW. To lowest order in the interaction and inhomogeneity, 
$a_s$ should have the form $a_s \teq a_0 - a_1 \sum_j |\bnabla\psi_j|^2$, where 
$a_0$ is the conventional quadratic coefficient ($a_0 \propto T - T_{0}$ if 
there are sources of pairing other than CDW fluctuations, which could lead to 
SC below some temperature $T_{0}$) and $a_1 \tgt 0$ so that SC is stabilized 
within regions of fluctuating 
C order (we take $a_1$ to be $T$-independent). This captures, 
phenomenologically, fluctuation-\emph{induced} ($a_0 \teq \text{const.}$) and 
fluctuation-\emph{enhanced} ($a_0 \tpropto T \tminus T_\text{sc}$) pairing, as 
well as the spatial enhancement of the electronic DOS at DCs \cite{Yan2017}.

The total free energy is now $\mathcal{F} \teq \mathcal{F}_{\text{cdw}} \tplus 
\mathcal{F}_{\text{sc}}$ and the coupling in \eqref{eq:f-sc} requires a 
self-consistent solution for both $\psi_j(\br)$ and $\Phi(\br)$. As in \TS{} 
$T_\text{cdw} \tsimeq 60$\,K and $T_\text{sc} \tsimeq 4$\,K${\ll\,} 
T_\text{cdw}$ \cite{Morosan2006,Kusmartseva2009,Li2015,Kogar2017}, 
the CDW is already well developed when SC emerges. This justifies solving the 
two problems independently, where $\mathcal{F}_\text{sc}$ is minimized subject 
to a passive CDW background $\psi_j(\br)$ determined by 
$\mathcal{F}_\text{cdw}$. (Although we note that the back-influence of a finite 
$\Phi(r)$ on $\psi_j(r)$ implied by \Eqref{eq:f-sc} increases CDW fluctuations 
via DCs so that SC and DCs mutually stabilize each other.)
A representative result \cite{SI} is shown in \Fref{fig:real-space}(d) for the 
CDW solution in panel (a)
\footnote{The SC boundary in \Fref{fig:phase-diagram} was obtained with $a_0 
\teq 10t+60$ and $b_s \teq c_s \teq 1$.}.
The most significant feature is the non-uniformity of $\Phi(\br)$ that follows 
the spatial texture of the DC network. The section plotted in 
\Fref{fig:real-space}(e) shows there is no SC within the C domains 
[$\Phi(\mathbf{x}_1) \teq 0$] but only at and near the DCs, and that SC is 
reinforced when two DCs overlap at the vertices of the Kagome: 
$\Phi(\mathbf{x}_3) \approx 2 \Phi(\mathbf{x}_2)$.

\begin{figure}
\centering
\includegraphics[width=0.5\textwidth ]{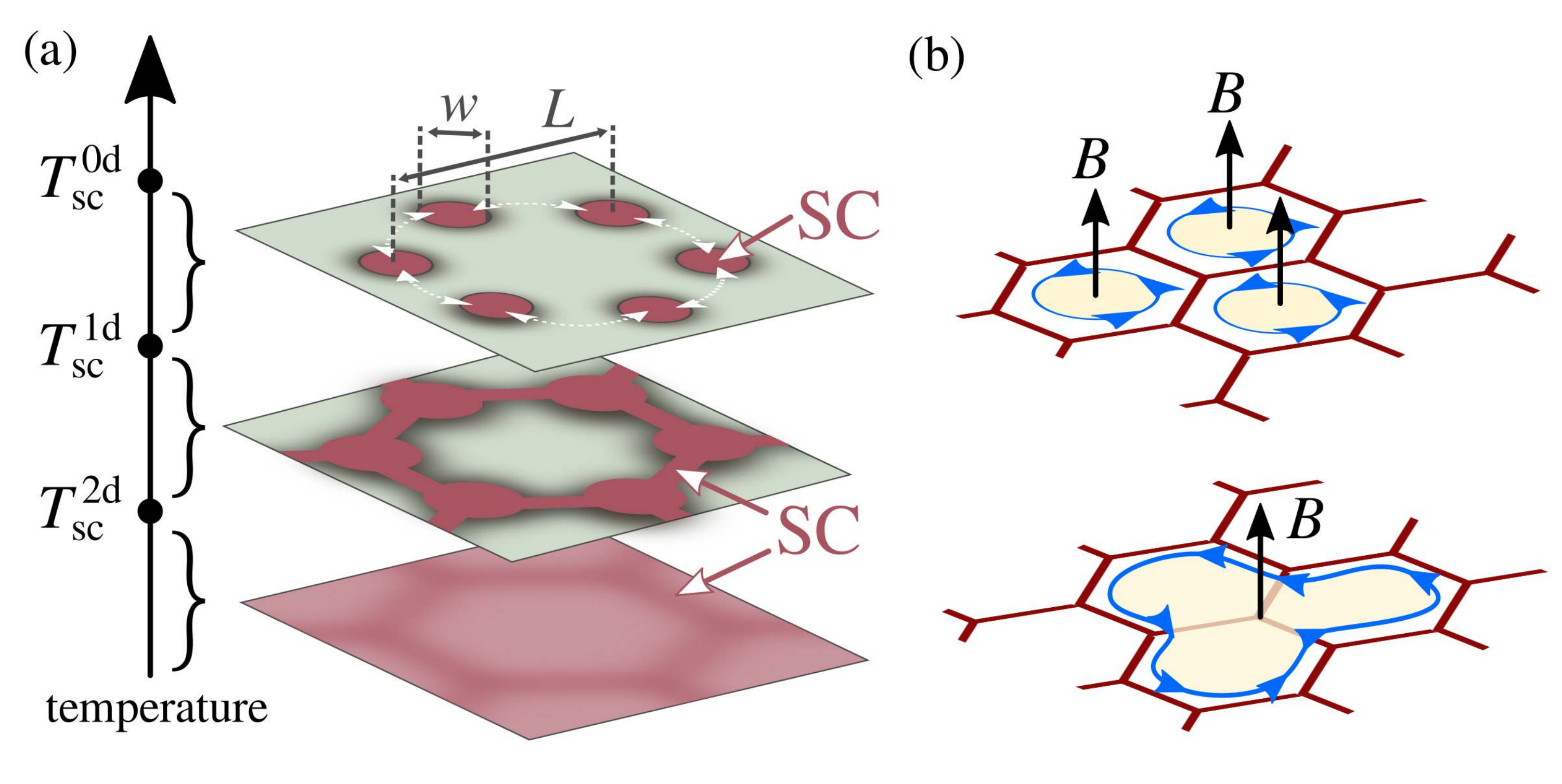}
\caption{
(a) Schematic of the distinct non-uniform SC regimes spatially correlated 
with the DC network: nucleation and expansion of the SC order parameter 
($T_\text{sc}^\text{1d} \tlt T \tle T_\text{sc}^\text{0d}$), percolation 
($T_\text{sc}^\text{2d} \tlt T \tle T_\text{sc}^\text{1d}$), and finite 
everywhere. See \SFref{4} for actually calculated textures.
(b) Illustration of how the connectivity in the percolation regime constrains 
the vortex structure, with impact in the magnetic response.
}
\label{fig:schematic}
\end{figure}

Interestingly, it is clear from how $\bnabla\psi_j$ enters the quadratic 
coefficient $a_s$ in \Eqref{eq:f-sc} that the development of SC in the NC regime 
can take place in three stages with decreasing temperature: (i) it begins at 
$T_\text{sc}^{\text{0d}}$ with the nucleation of isolated SC dots at the Kagome 
vertices, as illustrated at the top of \Fref{fig:schematic}(a) that depicts a 
unit cell of the DC/SC superlattice; (ii) at 
$T_\text{sc}^{\text{1d}} {\,\lesssim\,} T_\text{sc}^{\text{0d}}$ the dots 
have grown and overlap to percolate the system in a connected network as in 
\Fref{fig:real-space}(d); (iii) ultimately, at $T_\text{sc}^{\text{2d}} 
{\,\lesssim\,} T_\text{sc}^{\text{1d}}$ the whole system becomes 
superconducting. (The SC boundaries in the phase diagram correspond to 
$T_\text{sc}^{\text{0d}}$.)
The coupling proposed in \Eqref{eq:f-sc} therefore predicts that, 
depending on the temperature, the SC order can have either a 0d, 1d or 2d 
character. This can be directly probed with temperature-dependent local 
spectroscopy across the SC transition. In the absence of other pairing 
mechanisms, this picture predicts that if the penetration length of 
$\Phi(\br)$ into the C region is smaller than $L$, it is possible to have 
$T_\text{sc}^\text{2d} \teq 0$ in the NC region of the phase diagram. SC would 
then span the system, at most, through the 1d network defined by the DCs.

The area of SC stability in the phase diagram depends on whether the 
parameter $a_1$ in \Eqref{eq:f-sc} varies with $E$. If it does not, SC persists 
from the NC to the IC limit at temperatures below the gray line in 
\Fref{fig:phase-diagram}. It remains in the IC limit because $|\bnabla\psi_j|$ 
is finite, thereby supporting uniform SC. 
In the specific case of doped \TS{}, however, SC exists only in a dome-shaped 
portion of the phase diagram, over a finite density range 
\cite{Morosan2006,Li2015}. This phenomenology can be captured by replacing 
$a_1 \tto a_1 E$ in the parameter $a_s$, making it depend both implicitly 
(through $\psi_j$) and explicitly on the lock-in parameter $E$.
This amounts to making the coupling to CDW fluctuations weaker at higher 
densities, which is physically plausible in view of screening.
The SC boundary numerically recalculated in this way drops to lower
temperature when $E \rightarrow 0$, as conveyed by the red line in 
\Fref{fig:phase-diagram}, which qualitatively reproduces the experimental SC 
dome (see also Fig. S3).

% ------------------------------------------------------------------------------
% FIGURE
% ------------------------------------------------------------------------------
% ------------------------------------------------------------------------------

\textbf{Ramifications}\;---\;%
The feasibility of non-uniform percolative SC in the NC regime is determined 
by the characteristic width of DCs ($w$), their separation $L$ (the size 
of C domains), and the SC coherence length $\xi$ (${\sim}12$\,nm in 
\TS{} \cite{Morosan2007}). 
Likely, $w {\,\lesssim\,} \xi$, not sufficient to permit fully developed SC 
grains in the range $T^{\text{1d}}_\text{sc} \tlt T \tlt 
T^{\text{0d}}_\text{sc}$ where the model predicts nucleation at the vertices of 
the DC network.

The situation in the range $T^{\text{2d}}_\text{sc} \tlt T \tlt 
T^{\text{1d}}_\text{sc}$ has interesting implications in the presence 
of a magnetic field, $B$. First, vortices are naturally pinned by the DC 
lattice, even in the absence of disorder, and their motion correlated. 
Second, given the likelihood that $w{\,\lesssim\,}\xi$, vortices would not 
squeeze within DCs; the supercurrent would instead circulate along a linked 
network of 1D SC channels \cite{Tinkham2004}, as illustrated in 
\Fref{fig:schematic}(b). If $L{\,\gg\,}\xi$, we may regard this as a microscopic 
version of SC wire grids \cite{Pannetier84, Hallen1993, Ling1995, 
Stewart2007,Teitel83PRL, Alexander83, Niu89, Lin02, berger2001connectivity},
a distinctive feature of which are oscillatory dips as a function of $B$
in thermodynamic \cite{Pannetier84} and transport \cite{Ling1995} properties,
with period determined by rational fractions ($f\teq\phi/\phi_0$) of the  
flux through the grid's elementary plaquette ($\phi\sim B L^2$, $\phi_0 \tequiv 
h/2e$) \cite{Niu89, Park2001, Lin02}.

It is tempting to speculate whether such non-uniform SC texture can underlie 
the Little-Parks oscillations found in the SC phase of \TS{} near optimum 
doping \cite{Li2015}. To test it, assume the grid is hexagonal as in 
\Fref{fig:real-space}(d) ($f\teq1/4$ \cite{Lin02}) and take the first 
experimental magnetoresistance dip at $B\tsimeq 0.13$\,T.
With our results, we obtain the incommensurability factor $\delta\tsim 
0.01$ and a typical distance between DCs $L\tsim70$\,nm 
\footnote{Since $a\teq 0.35$\,nm for \TS{} and $\eta\tsim 1$ 
(\Fref{fig:phase-diagram}), we have $L \teq\sqrt{3}a/(\eta \delta) \teq 
[\phi_0/(2\sqrt{3}B)]^{1/2} \tsimeq 70$\,nm and $\delta\tsim 0.01$.}.
Compellingly, x-ray diffraction does reveal $\delta\tsim$5--15\% in the 
superconducting dome \cite{Kogar2017}, and STM finds DCs separated by 10's of 
nm at optimum doping above $T_\text{sc}$ \cite{Yan2017}. It is noteworthy how 
these estimates agree with experiments.

Our model captures qualitatively well the emergence of SC correlated with the 
suppression of the C-CDW. This phenomenology is not unique to \TS{}, but 
documented across a number of $2H$ and $1T$ TMDs \cite{Wang2017} spanning both 
good metals and semimetals, as well as distinct commensurability 
conditions. Our approach straightforwardly extends to those cases \cite{SI}, 
providing a definite and universal phenomenological foundation to further 
explore the interplay between these two coexisting orders and their 
fluctuations.

\begin{acknowledgments}
VMP was supported by the Ministry of Education of Singapore through grant 
MOE2015-T2-2-059 and AHCN by the National Research Foundation of Singapore under 
its Medium-Sized Centre Programme. Numerical computations were carried out at 
the HPC facilities of the NUS Centre for Advanced 2D Materials.
\end{acknowledgments}

\bibliographystyle{apsrev4-1}
\bibliography{G-L-theory}

% ------------------------------------------------------------------------------
% SUPPLEMENTARY INFORMATION
% ------------------------------------------------------------------------------

% Note: I've only been able to avoid page superposition in the attached pdf 
% below by explicitly adding an empty page between each by hand. 
% See https://tex.stackexchange.com/questions/15989

\clearpage
\includepdf[pages={{},1,{},2,{},3,{},4,{},5,{},6,{},7,{},8,{},9,{},10,{},11,{},12,{},13,{},14,{},15,
{},16,{},17,{},18,{},19,{},20,{},21,{},22,{},23}]%
{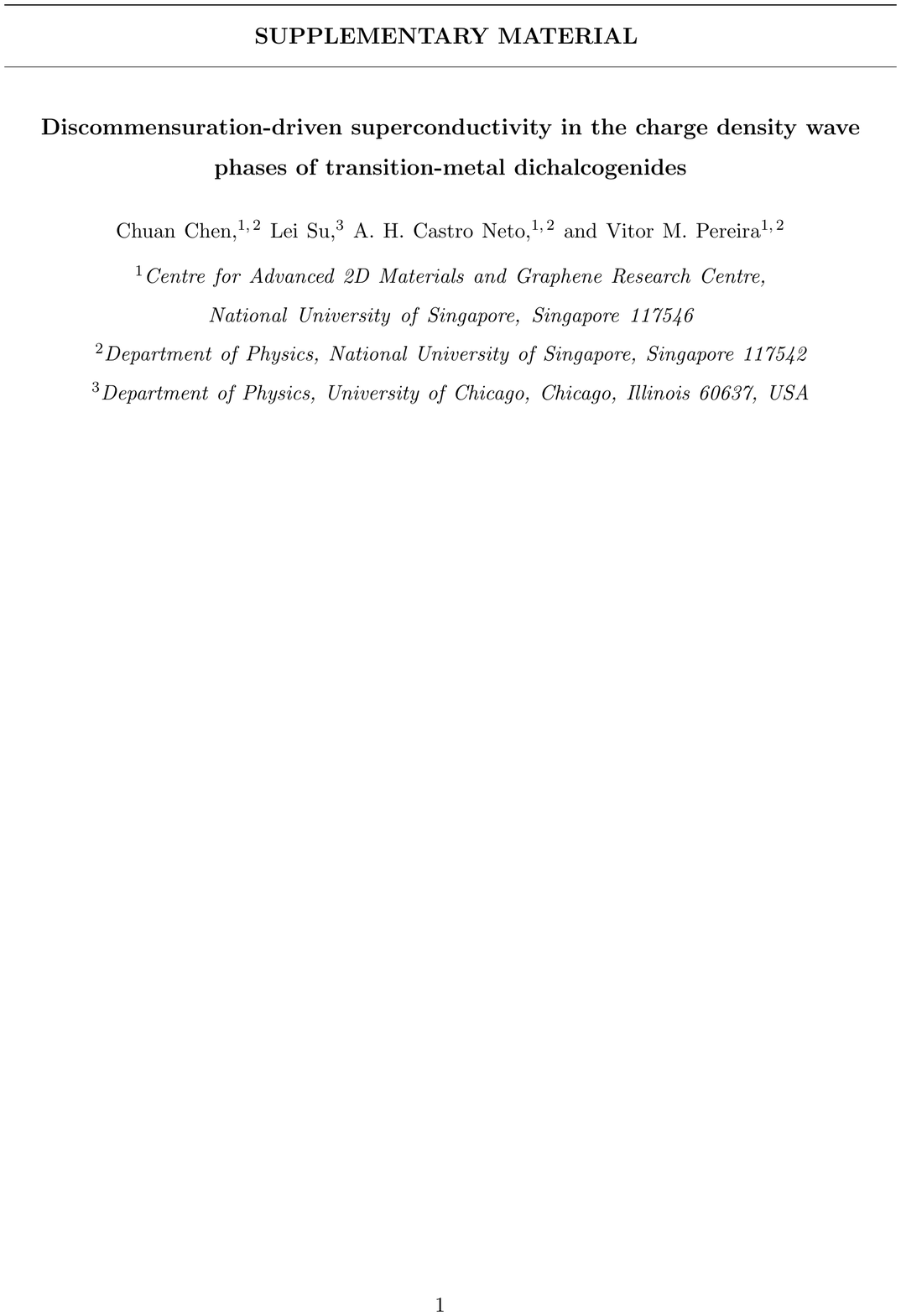}

\end{document}